# Ferroelectric switching of a two-dimensional metal

Zaiyao Fei[1†], Wenjin Zhao[1†], Tauno A. Palomaki[1†], Bosong Sun[1], Moira K. Miller[1], Zhiying Zhao[2], Jiaqiang Yan[2], Xiaodong Xu[1,3], and David H. Cobden[1]*

[1]Department of Physics, University of Washington, Seattle, WA 98195, USA
[2] Materials Science and Technology Division, Oak Ridge National Laboratory, Oak Ridge, Tennessee 37831, USA
[3]Department of Materials Science and Engineering, University of Washington, Seattle, WA 98195, USA
[†] These authors contributed equally    *Corresponding author: cobden@uw.edu

**A ferroelectric is a material with a polar structure whose polarity can be reversed by applying an electric field[1,2]. In metals, the itinerant electrons tend to screen electrostatic forces between ions, helping to explain why polar metals are very rare[3-7]. Screening also excludes external electric fields, apparently ruling out the possibility of polarity reversal and thus ferroelectric switching. In principle, however, a thin enough polar metal could be penetrated by an electric field sufficiently to be switched. Here we show that the layered topological semimetal $WTe_2$ provides the first embodiment of this principle. Although monolayer $WTe_2$ is centrosymmetric and thus nonpolar, the stacked bulk structure is polar. We find that two- or three-layer $WTe_2$ exhibits a spontaneous out-of-plane electric polarization which can be switched using gate electrodes. We directly detect and quantify the polarization using graphene as an electric field sensor[8]. Moreover, the polarization states can be differentiated by conductivity, and the carrier density can be varied to modify the properties. The critical temperature is above 350 K, and even when $WTe_2$ is sandwiched in graphene it retains its switching capability at room temperature, demonstrating a robustness suitable for applications in combination with other two-dimensional materials[9-12].**

The structure of a polar material effectively contains an arrow, that is, an axis along which the two directions are physically distinguishable. This is necessary for the existence of a spontaneous electric polarization. Of the 32 three-dimensional crystal classes, the ten having a polar axis are known as the pyroelectrics, because heating them changes any electric polarization along this axis to produce a voltage. When Anderson and Blount introduced[3] the term ferroelectric metal in 1965, they were referring to the possibility of polar structure appearing in certain metallic crystals on cooling. They naturally assumed that, even if such polar metals exist, the polarity would not be switchable. The first definite cases of metals with polar structure have only very recently been identified[4-7].

Several ferroelectric insulators have been found to maintain ferroelectric characteristics in ultrathin films[13-17]. However, when materials with a layered structure are thinned down towards the monolayer limit their properties often change qualitatively. This is amply illustrated by graphene, which becomes a two-dimensional Dirac metal[11,12]; by $MoS_2$, which changes from indirect to direct gap semiconductor[10]; and by $CrI_3$, which varies between antiferromagnetic and ferromagnetic[9]. Another example is the topological semimetal $WTe_2$[18], which becomes either a two-dimensional topological insulator[19-22] or a superconductor at low temperatures in the monolayer limit, depending on the electrostatic doping level. Here we focus on another aspect of $WTe_2$: it is a polar metal. The three-dimensional (1T′) structure has a polar space group[7], *Pnm2₁*, and it remains metallic down to three layers when undoped[23] or even as a monolayer when electrostatically doped[20]. We show here that as it approaches this limit the polarity can be switched, making it effectively ferroelectric even while it is metallic in the plane.

The 1T′ structure (Fig. 1a) contains *b-c* mirror (M) and *a-c* glide (G) planes, so that the polar axis, which must be parallel to both of them, is the *c*-axis, perpendicular to the layers[7,18]. We can apply an electric field along this axis using the device geometry indicated in Fig. 1b. An electrically contacted thin $WTe_2$ flake is sandwiched between two hexagonal boron nitride (hBN) dielectric sheets, with thicknesses $d_t$ (top) and $d_b$ (bottom). Above and below are gate electrodes, usually of few-layer graphene, to which voltages $V_t$ and $V_b$ are applied relative to the grounded $WTe_2$ (see Methods, and Extended Data Fig. 1 and Table 1, for device fabrication and characterization).

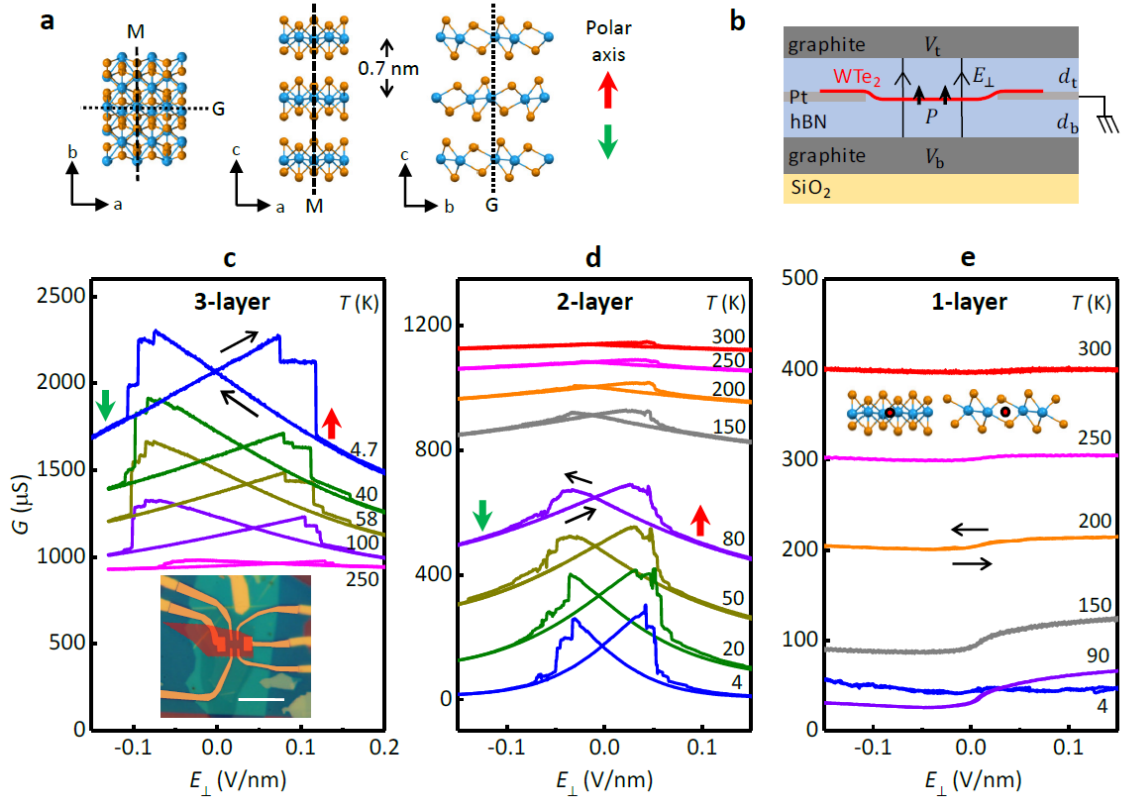

**Figure 1. Evidence for ferroelectric switching in WTe₂. a,** Structure of three-dimensional 1T′ WTe₂, showing mirror (M) plane, glide (G) plane and polar *c*-axis. **b,** Schematic cross-section of the device geometry used to apply electric field $E_\perp$ normal to an atomically thin WTe₂ flake. **c, d,** Conductance of undoped trilayer device T1 and bilayer device B1 as $E_\perp$ is swept up and down, setting $V_t/d_t = -V_b/d_b$ to avoid net doping. It shows bistability associated with electric polarization up (red arrow) or down (green arrow), at temperatures from 4 K to 300 K. Here the conductance is the reciprocal of the four-terminal resistance. The undoped trilayer has a metallic temperature dependence, the bilayer an insulating one. Inset to **c**: optical image of a representative double-gated device. The WTe₂ flake has been artificially colored red; the scale bar is 10 μm. **e,** Similar measurements on a monolayer WTe₂ device (M1), showing no bistability. At 4 K conduction is in the quantum spin Hall (QSH) regime. Insets: location (red dot) of a center of symmetry in the monolayer, viewed along *b* and *a* axes.

We define $E_\perp = (-V_t/d_t + V_b/d_b)/2$, the applied electric field passing upwards through the layer, which will couple to out-of-plane polarization. When $E_\perp$ is swept up and down the conductance of 3-layer (Fig. 1c) and 2-layer (Fig. 1d) devices shows bistability near $E_\perp = 0$ characteristic of ferroelectric switching, at all temperatures *T* from 1.6 K up to above room temperature. No bistability is seen in monolayer WTe₂ (Fig. 1e), consistent with its structure having a center of symmetry (Fig. 1e inset, red dots) and hence being non-polar; this also rules out instabilities involving charge injection into the hBN. Nor is bistability seen in thicker crystals, including in the situation where one is employed as a gate electrode (see Extended Data Fig. 2). This, and the larger field required to switch the trilayer than the bilayer, can be explained by screening of $E_\perp$ on a length scale of nanometers.

Similar bistability was seen in all bilayer devices (see Extended Data Fig. 3). To prove that it is associated with out-of-plane electric polarization, we made devices in which the top gate is replaced by monolayer graphene, whose conductivity is sensitive to the precise electric field $E_t$ in the upper hBN.

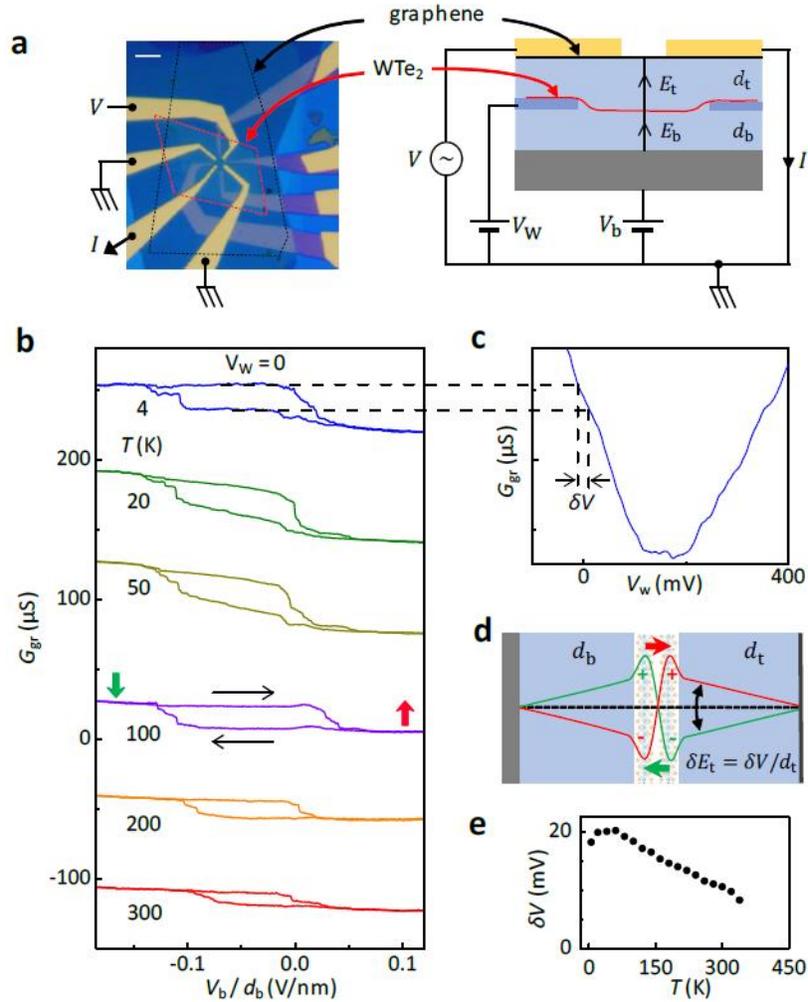

**Figure 2. Detecting the out-of-plane polarization. a,** Micrograph and schematic cross-section of a bilayer WTe$_2$ device (B2) having multiply contacted graphene in place of the top gate, indicating separately the electric fields in the hBN above ($E_t$) and below ($E_b$) the WTe$_2$. Scale bar: 5 μm. **b,** The graphene conductance $G_{gr}$ is measured when a bias $V_b$ is applied to the bottom gate with the intervening WTe$_2$ grounded, at a series of temperatures. The two conductance states seen for the two sweep directions are associated with different out-of-plane polarization states of the WTe$_2$ (red and green arrows). **c,** The behavior of $G_{gr}$ (the y-axis is the same as **b**) when a voltage $V_W$ is applied directly to the WTe$_2$ provides a mapping to the difference $\delta E_t = \delta V/d_t$ in $E_t$ between the two states. **d,** Sketch indicating how the reversal of the polarization changes the electrostatic potential (from red to green) and $E_t$ (see text). **e,** Temperature dependence of $\delta V$, which is proportional to the polarization.

Figure 2 presents measurements at a series of temperatures on such a bilayer WTe$_2$ device (B2) with four gold contacts to the top graphene, depicted in Fig. 2a (see Extended Data Fig. 4). If the WTe$_2$ acts as a conducting sheet it will screen out any electric field due to a voltage applied to the bottom gate. Indeed, Fig. 2b shows that the conductance $G_{gr}$ of the graphene depends only very weakly on $V_b$, except in a certain interval where it jumps between two states. The conductance of the WTe$_2$ is bistable in precisely the same interval (see Extended Data Fig. 4). The two states must correspond to different values of $E_t$ that can occur for exactly the same set of applied bias voltages. This implies the existence of two different vertical distributions of charge in the bilayer WTe$_2$. We deduce that sweeping the bottom gate changes $E_\perp$ (here $E_\perp = V_b/2d_b$ since $V_t = 0$), which at the ends of the hysteresis loop flips the polarization state (henceforth denoted by P↑ or P↓), changing $E_t$ by an amount $\delta E_t$, and so changing $G_{gr}$.

We can infer $\delta E_t$ by applying a bias $V_W$ to the WTe$_2$ and measuring the change $\delta V = d_t \delta E_t$ required to produce the same change in $G_{gr}$, as shown in Fig. 2c. For the simplified case $d_t = d_b$ and all voltages at zero, the electrostatic potential profile is inverted between P↑ (red) and P↓ (green), as sketched in Fig. 2d, and the areal polarization density is given by $P \approx \epsilon_0 \delta V$ (see Methods). At 20 K this gives $P \approx 1 \times 10^4$ $e \cdot$cm$^{-1}$, equivalent to transferring $\sim 2 \times 10^{11}$ $e \cdot$cm$^{-2}$ between the two layers, a distance of ~0.7 nm. This is three orders of magnitude lower than the volume polarization density ~0.2 C·m$^{-2}$ ~ $10^{14}$ $e \cdot$ cm$^{-2}$ in the classic ferroelectric[1] BaTiO$_3$. Combined with the micron-scale device size, such a small polarization makes it very hard to detect the ferroelectricity by standard displacement current measurements.

In Fig. 2e we plot $\delta V$ as a function of temperature. Between ~60 K and 300 K it decreases roughly linearly with $T$, extrapolating to zero at roughly 450 K. However, above ~340 K the signal becomes unstable and we can no longer identify a hysteresis loop, suggesting that a transition to a nonpolar state occurs in this temperature range.

We also made a simpler device with no top hBN and monolayer graphene directly encapsulating bilayer WTe$_2$. It exhibited highly reproducible hysteresis in the conductance, visible up to 300 K (see Extended Data Fig. 5), showing that the ferroelectric switching is robust enough for applications in combination with other 2D materials at room temperature.

We also investigated the effect of the gate-induced charge doping, defined by $en_e = \epsilon_{hBN}\epsilon_0(V_t/d_t + V_b/d_b)$. The quantity $n_e$ would be the areal density of added electrons if the material were a simple metal. In Figs. 3a and b we plot the conductance at 7 K for device B1 (the same bilayer as in Fig. 1d), as a joint function of $V_t$ and $V_b$, measured with $V_t$ stepped and $V_b$ swept up or down. Each sweep was started in the same fully polarized state. The black dashed line denotes $E_\perp = 0$ and the white dashed line $n_e = 0$. The two plots differ only in the central hysteretic region, as is made clearer by plotting the difference between them (Fig. 3c). Similar behavior is seen at higher temperature (Fig. 3d, at 200 K). At $E_\perp = 0$, $G$ is a similar function of $n_e$ for both P↑ and P↓ (Fig. 3e), with temperature dependence that is insulating near $n_e = 0$ and metallic for $n_e > n_c \sim 2 \times 10^{12}$ cm$^{-2}$, as reported previously[20]. Figure 3f shows traces obtained by sweeping $E_\perp$ repeatedly up and down for selected values of $n_e$ at 7 K. In each case the single conductance level at large $E_\perp$ evolves smoothly and reproducibly into one of the two stable levels as $E_\perp$ is reduced to zero, implying that the state remains uniformly polarized, without domain structure, at $E_\perp = 0$. For small or negative $n_e$, the effect of $E_\perp$ is large and of opposite sign for P↑ and P↓, producing butterfly-shaped hysteresis loops. For $n_e$ well above $n_c$, $E_\perp$ has less effect on the conductance and the hysteresis is smaller but it is still present. Hence the doped bilayer, like the trilayer, is simultaneously ferroelectric and metallic.

At low temperatures (Fig. 3c) we observe an increase of the hysteresis loop width for increasingly negative $n_e$, while at 200 K (Fig. 3d) it is almost independent of $n_e$. When the conductance jumps there is some stochastic variation in the positions and substructure of the jumps suggesting domain dynamics. If the surrounding gates were not present to screen the depolarization field, domains would inevitably form to limit the electrostatic energy[24], as observed in other ultrathin ferroelectrics[13,16,25,26]. In our devices, defects such as rips, bubbles and folds could nucleate domains or pin domain walls. In addition, $E_\perp$ is not completely uniform, because above and near the Pt contacts it will be reduced by screening. Indeed, the pattern of switching depends on choice of measurement contacts within a given device (see Extended Data Fig. 6). We also observe that in some bilayer devices, such as B2 (see Fig. 2), the switching field is not symmetric about $E_\perp = 0$. A possible explanation for this is that sometimes, despite all precautions, during device fabrication one side of the WTe$_2$ flake is exposed to mild oxidation, producing asymmetric trapped charge.

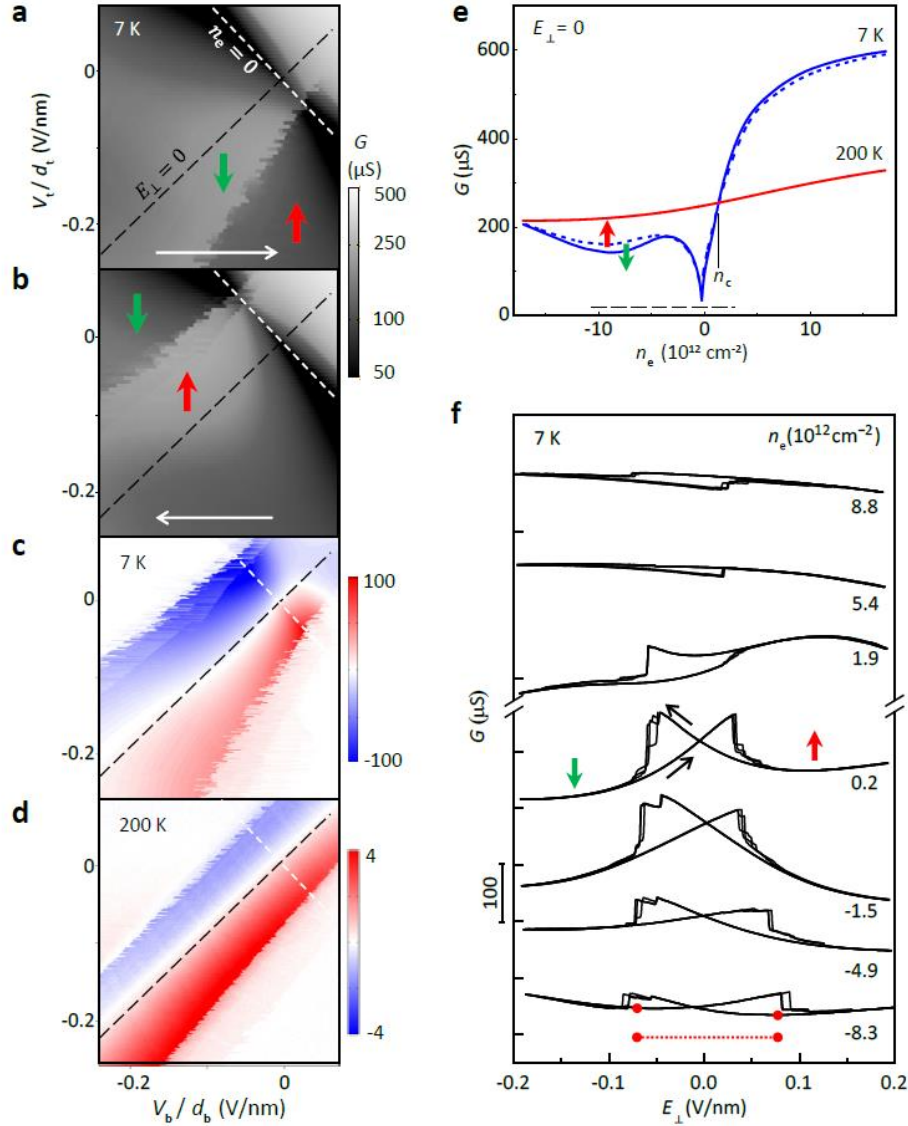

**Figure 3. Gate tuning of the ferroelectric behavior. a, b,** Conductance $G$ of bilayer device B1 at 7 K as a function of both gate voltages, for the two sweep directions of $V_b$ as indicated. **c,** Difference between the above two plots at 7 K, which is nonzero in the hysteretic regime. **d,** Same measurement at 200 K. Dashed lines indicate contours of zero perpendicular field $E_\perp$ and zero charge density $n_e$. **e,** variation of $G$ with $n_e$ at $E_\perp = 0$ for both polarization states at two temperatures. The dashed bar indicates the range of $n_e$ in **a-d**. **f,** Sweeps of $E_\perp$ for fixed $n_e$ at 7 K. The dotted bar near the bottom indicates the magnitude of the approximate shift in $E_\perp$ of the conductance minimum between the opposite polarization states.

The fact that the conductance is sensitive to the polarization is consistent with the expectation that the polarization redistributes charge between the layers, which are inequivalent when $E_\perp$ is nonzero. While the specific mechanisms for the sensitivity to $n_e$, $E_\perp$, and $P$ are still under investigation, a few remarks are in order. First, the monolayer conductance at 4 K in Fig. 1e, which is due to edge conduction since this is the established QSH regime, is almost independent of $E_\perp$. Second, in bilayers at large positive or negative $n_e$ the reversal of $P$ has a similar effect on the conductance to changing $E_\perp$, by an amount ~0.15 V/nm at 7 K (indicated by dotted horizontal line in Fig. 3f). This corresponds to a change in the electrostatic potential difference between the two WTe$_2$ layers by ~100 mV. This is of the same order as the estimated change in the potential difference associated with the polarization reverse, $2\delta V \sim 40$ mV, suggesting that the potential imbalance between the layers governs the sensitivity of the conductance to both $E_\perp$ and $P$. It is also roughly the same as the width of the hysteresis loop; that is, the polarization flips roughly when the applied potential difference exceeds the potential due to the spontaneous polarization. This is another indicator that electron transfer between the layers may be involved. Third, the very sharp minimum seen in $G$ close to $n_e = 0$ in bilayers, visible in Fig. 3e, presumably marks the compensation point where electron and hole densities are exactly equal, suggesting that electron-hole correlation may be important. Taken together, the above observations raise the intriguing possibility that electron-hole correlation effects, rather than a lattice instability[27], drive the spontaneous polarization in this layered semimetal. If this is the case, the polarization could principally involve a relative motion of the electron cloud relative to the ion cores, rather than a lattice distortion, in which case the switching would be intrinsically very fast.

Ferroelectricity adds another ingredient to the intriguing combination of QSH edges, correlation effects and superconductivity already seen in this atomically thin topological semimetal. Although the QSH behavior and superconductivity are restricted to the centrosymmetric monolayer while the ferroelectricity occurs only for two or more layers, it is possible that these diverse phenomena are connected in ways which may also be relevant to understanding the properties that emerge in the 3D limit, including extreme and anisotropic magnetoresistance[28,29], a polar axis, and Weyl points[18,30].

**Acknowledgements**
We thank Joshua Folk, Ebrahim Sajadi, Arkady Levanyuk, Turan Birol and Anton Andreev for significant insights. DHC and XX were supported by the U.S. Department of Energy, Office of Basic Energy Sciences, Division of Materials Sciences and Engineering, under Awards DE-SC0002197 and DE-SC0018171 respectively. Synthesis efforts at ORNL (ZZ and JY) were also supported by the same division of the DoE. TP was supported by AFOSR FA9550-14-1-0277. ZF, WZ and BS were supported by the above awards and also by NSF EFRI 2DARE 1433496 and NSF MRSEC 1719797.

**Methods**

**Preparation and characterization of WTe$_2$ devices.** We measured devices with four different layouts: (1) WTe$_2$ with graphite gates above and below (M1, B1, B4, T1); (2) bilayer WTe$_2$ with monolayer graphene as top gate (B2); (3) a bilayer WTe$_2$/graphene heterostructure (B3); and (4) a monolayer graphene device gated by few-layer WTe$_2$ (F1). In the following, we describe fabrication of the first type; the others are similar.

First, graphite and hBN crystals were mechanically exfoliated under ambient conditions onto substrates consisting of 285 nm thermal SiO$_2$ on highly p-doped silicon. Graphite flakes 2-6 nm thick were chosen for top and bottom gates and 5-30 nm thick hBN flakes (a layered electrical insulator free of trapped charges and dangling bonds) were chosen for the top and bottom dielectric[31]. The top and bottom parts were prepared separately using a polymer-based dry transfer technique[32]. For the bottom part, an hBN flake was picked up on a polymer stamp and placed on the bottom graphite. After dissolving the polymer, Pt metal contacts (~8 nm) were patterned on the hBN by standard e-beam lithography, e-beam evaporation and lift-off. For the top part, the top graphite was picked up first, then the top hBN. Both stacks were then transferred to an oxygen/water-free glovebox. WTe$_2$ crystals were exfoliated inside the glovebox and flakes from monolayer to trilayer thickness were optically identified and quickly picked up with the top part; the stack was then completed by transferring onto the lower contacts/hBN/graphite stack before taking out of the glovebox. Finally, after dissolving the polymer, another step of e-beam lithography and metallization was used to define electrical bonding pads (Au/V) connecting to the metal contacts and the top and bottom gates. Extended Data Fig. 1 shows schematics of the fabrication processes, optical and AFM images of a typical bilayer WTe$_2$ device (B4).

**Estimation of the electric polarization.** We use the following simplified model to estimate the spontaneous polarization of the bilayer WTe$_2$ from the measurements in Fig. 2b. We assume that $d_\text{t} = d_\text{b} \gg d$, where $d$ is the thickness of the WTe$_2$; that all conductors (bottom graphite gate, top graphene and bilayer WTe$_2$) are grounded and have infinite electronic compressibility; and that the areal polarization density $P$ is associated with two thin sheets of areal charge density $\pm P/d$ separated by $d$. Under these assumptions, when the polarization reverses there is no net flow of charge between the conductors and the WTe$_2$ remains neutral, and the potential profile between the gates is simply reversed when the polarization flips (Fig. 2d). By Gauss's law,

$$\epsilon_0 \epsilon_\text{hBN} E_\text{t} = \epsilon_0 E_\text{i} + P/d , \qquad (1)$$

where $E_t$ is the electric field in the hBN (equal on both sides because the bilayer is neutral) and $E_\text{i}$ is the field between the two charge sheets. Since the top graphene and the center of the bilayer are both at zero potential,

$$2E_\text{t} d_\text{t} + E_\text{i} d = 0 . \qquad (2)$$

From (1) and (2),

$$E_t = \frac{P}{\epsilon_0 (2d_\text{t} + \epsilon_\text{hBN} d)} . \qquad (3)$$

The change in $E_t$ when the polarization reverses is then $\delta E_t = 2E_t = \frac{2P}{\epsilon_0(2d_t+\epsilon_{hBN}d)}$. With $d_t \approx 10$ nm and $d \approx 1$ nm, the first term in the denominator dominates so $\delta E_t \approx P/(\epsilon_0 d_t)$ and thus

$$P \approx \epsilon_0 d_t \delta E_t = \epsilon_0 \delta V \ .$$

In reality, the ratio of $d_t$ to $d_b$ could be up to about 1:3, the conductors have finite compressibility, and the polarization charge is more spread out, which taken together will introduce an extra numerical coefficient of order unity.

**Removing parallel (parasitic) conduction through the graphene in device B2.** In device B2 the graphene extends over regions with no WTe$_2$ underneath so that it acts as a uniform gate for the entire WTe$_2$ sheet. The quantity that we call $G_{gr}$ is the result of the following measurement, which maximizes sensitivity to just a central region of graphene above the WTe$_2$. First, we ground two opposing contacts to the graphene and measure only the current that flows from the biased contact to the one opposite, as shown in Fig. 2a. However, because of finite contact resistance, a small portion of this current still flows through graphene not above the WTe$_2$. To remove this parasitic current component, we set the WTe$_2$ voltage $V_W$ such that the graphene is at its Dirac-point minimum in the region over the WTe$_2$. Since the minimum is quite broad, the graphene over the WTe$_2$ is then insensitive to $V_b$ and the measured dependence on $V_b$ comes only from the parasitic component which can then be subtracted out. Note that removing it has no effect on the magnitude of the hysteresis.

In Extended Data Fig. 4b, we illustrate this procedure at 220 K. From the inset of Extended Data Fig. 4b, we determine that the graphene above the WTe$_2$ is at its Dirac point at $V_W = 129$ mV. The red curve shows the conductance of the graphene, $G_{gr}$, when $V_W = 129$ mV; the dependence on the back gate is from only the parasitic contribution. Conversely, in Fig. 2b and the blue curve in Fig. S4b we measure $G_{gr}$ at $V_W = 0$ mV, where the graphene is most sensitive to changes in the electric field in the top BN, $E_t$, yet also contains the parasitic conductance. The difference between these two curves (at $V_W = 129$ mV and 0 mV) is shown in black. The hysteresis remains, while the "V" shape is mostly removed. The remaining small slope can be explained by the finiteness of the electronic compressibility of the bilayer WTe$_2$.

Using Extended Data Fig. 4b we can estimate the ratio of the parasitic current to that flowing above the WTe$_2$. The area with no WTe$_2$ has a BN thickness of $d_t + d_b = 33$ nm between the graphene and bottom gate. The red curve (with a parasitic $V_b$ dependence) has a maximum slope of $dG_{gr}/dV_b = 17$ μS/V or $dG_{gr}/dE_t \sim 560$ μS/(Vnm$^{-1}$) after taking into account the BN thickness. From the inset curve, using voltage $V_W$ applied to the WTe$_2$ for gating (with 8 nm BN) gives $dG_{gr}/dE_t = 12300$ μS/(Vnm$^{-1}$). Thus the parasitic component is only ~5% of the total current.

**Methods references**

# Supplementary Information

**Content:**

**1. Extended Data Figures 1-6**
**2. Extended Data Table 1**

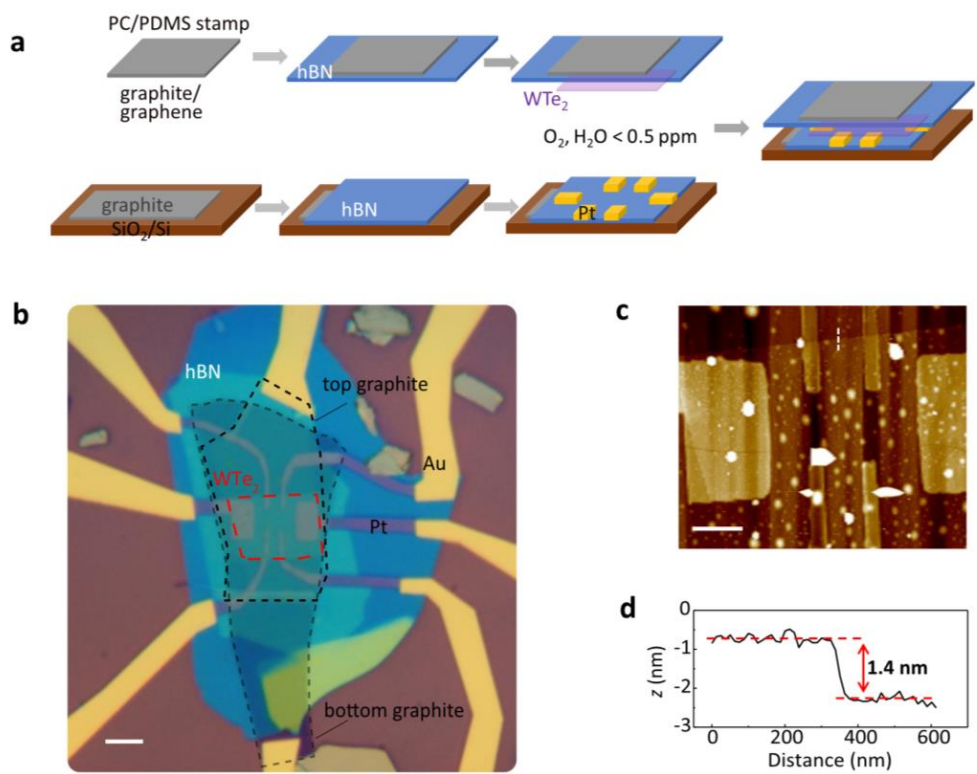

**Extended Data Figure 1: Bilayer WTe$_2$ device. a,** Essential steps in device fabrication. **b,** Optical image of device B4. The red dashed line outlines the bilayer flake. Scale bar: 5 μm. **c,** Atomic force microscope topography image of the central region in (a). Scale bar: 2 μm. **d,** Line cut along the white dashed line in (b). The step height matches the expected bilayer thickness, ~1.4 nm.

| Device label | WTe₂ | top hBN (nm) | bottom hBN (nm) | $C_t$ ($1\times10^{-3}$ F/m²) | $C_b$ ($1\times10^{-3}$ F/m²) |
|---|---|---|---|---|---|
| M1 | monolayer | 6 | 28 | 5.9 | 1.3 |
| B1 | bilayer | 12 | 20 | 3.0 | 1.8 |
| B2 | bilayer | 8 | 25 | 4.4 | 1.4 |
| B3 | bilayer | NA | 24 | NA | 1.5 |
| B4 | bilayer | 10 | 21 | 3.5 | 1.7 |
| T1 | trilayer | 5.5 | 23 | 6.4 | 1.1* |
| F1 | 8 nm | 24 | NA | 1.5 | NA |

**Extended Data Table 1: Thickness of hBN dielectrics and corresponding areal capacitances for the WTe₂ devices.** We define the gate-induced density imbalance to be $n_e = (C_t V_t + C_b V_b)/e$, and $E_\perp = (-C_t V_t + C_b V_b)/2\epsilon_{hBN}\epsilon_0$, where the geometric areal capacitances are $C_t = \epsilon_{hBN}\epsilon_0/d_t$, $C_b = \epsilon_{hBN}\epsilon_0/d_b$, $\epsilon_{hBN} \approx 4$ is the dielectric constant of hBN, $d_t$ and $d_b$ are the thicknesses of the top and bottom hBN flakes, respectively. All thicknesses were obtained from AFM images. In device B3 the WTe₂ flake is directly under the top graphene (no top hBN). In device F1 the thick WTe₂ is directly on the bottom graphite (no bottom hBN). *For device T1, there is no bottom graphite; instead, we used the metallic silicon substrate as the bottom gate, the areal capacitance then being $C_b = \epsilon_0 \left( d_b/\epsilon_{hBN} + d_{SiO_2}/\epsilon_{SiO_2} \right)^{-1}$.

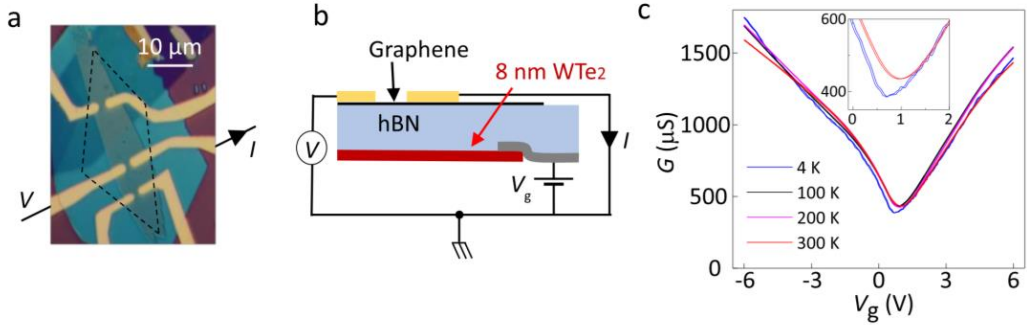

**Extended Data Figure 2: Thick WTe₂ used as a gate. a,** Optical image of device F1, in which a thick (8 nm) WTe₂ flake under 24 nm hBN is used as a gate for a top graphene sheet. Scale bar: 10 μm. **b,** Schematic cross-section of the device. **c,** Two-terminal conductance of the graphene as a function of voltage $V_g$ applied to the WTe₂ flake. There is no sign of switching or bistability at any temperature, indicating that no polarization reversal occurs on the WTe₂ surface for fields of up to $E_\perp \approx 0.125$ V/nm. Inset: zooms on the graphene Dirac point at 4 K and 300 K.

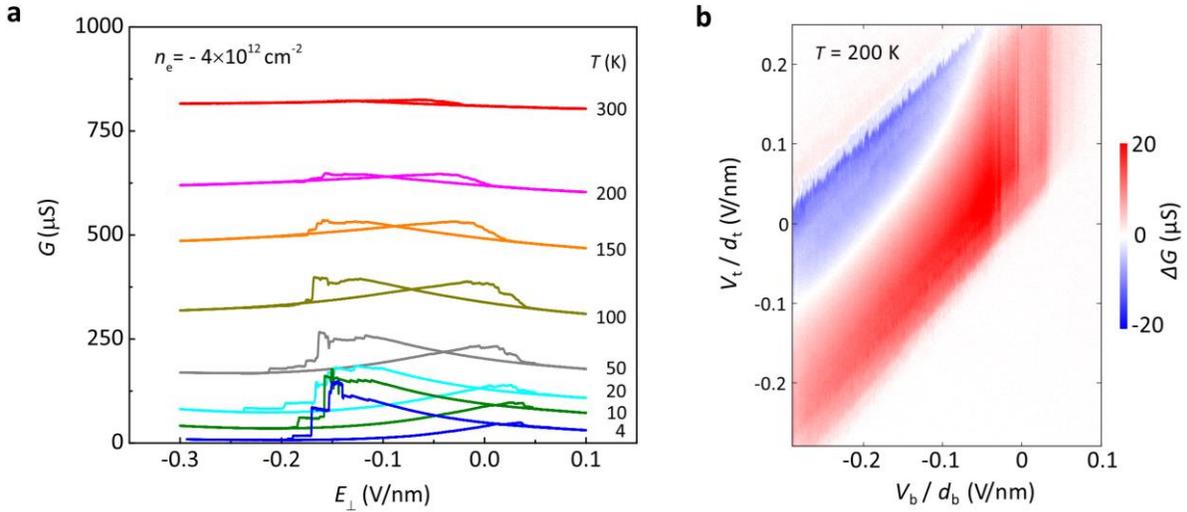

**Extended Data Figure 3: Switching of an additional bilayer device. a,** Conductance vs perpendicular electric field $E_\perp$, at temperatures from 4 K to 300 K, and gate doping level $n_e = -4 \times 10^{12}$ cm$^{-2}$ for device B4. **b,** Conductance difference between the two sweep directions of $V_b$ at 200 K, as plotted in Fig. 3d for device B1.

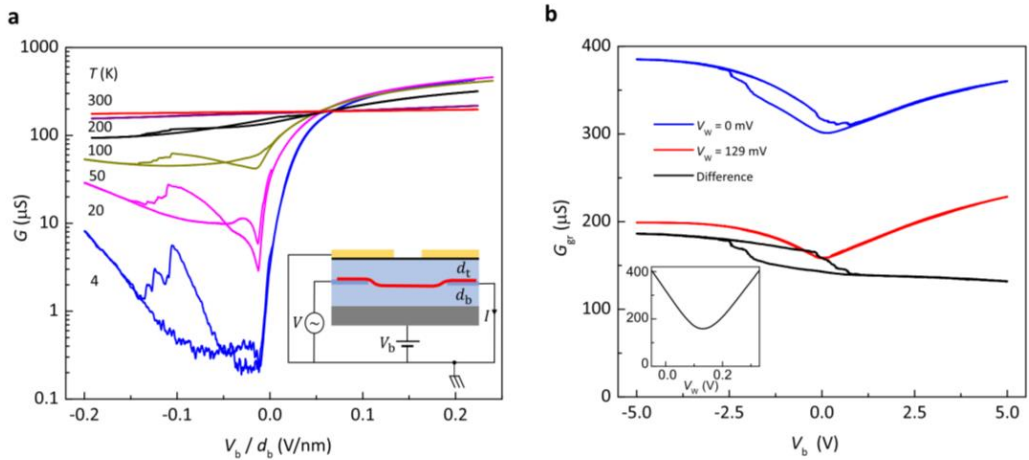

**Extended Data Figure 4 | Additional transport measurements and removal of parasitic effects in the polarization measurements. a,** Conductance vs $V_b$ for the bilayer WTe$_2$ in device B2, measured with the top graphene grounded. The hysteresis occurs in exactly the same range of $E_\perp$ as it does in the graphene conductance in Fig. 2b. Note that both $n_e$ and $E_\perp$ change when $V_b$ is swept. **b,** Graphene conductance at 220 K as a function of $V_b$ with the the voltage $V_W$ on the bilayer WTe$_2$ at 0 mV (blue) and 129 mV (red) respectively. The black curve is the difference between the blue and red curves. This removes most of the $V_b$ dependence of the "parasitic" current that flows through the top graphene which is not screened from the bottom gate by the WTe$_2$. Inset: Graphene conductance showing the minimum at $V_W = 129$ mV.

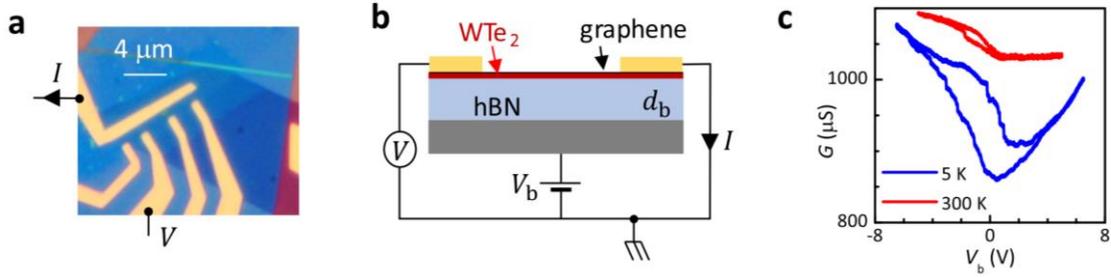

**Extended Data Figure 5 | Graphene/bilayer WTe₂ heterostructure showing hysteresis up to room temperature. a,** Device image, and **b,** schematic cross-section. **c,** The two-terminal conductance showing bistability at both 5 K and room temperature, implying that the polarization of the WTe₂ is still present in this hybrid structure.

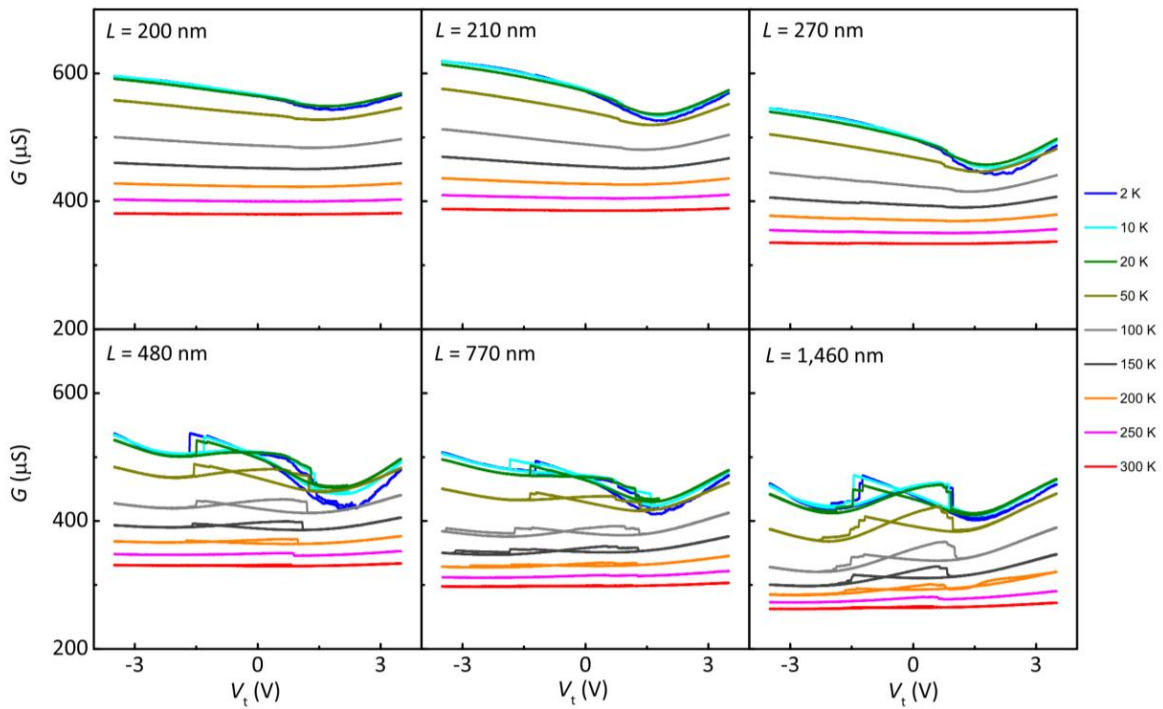

**Extended Data Figure 6| Length dependent ferroelectric behavior in trilayer WTe₂ for temperature from 2 K to 300 K.** All measurements are performed at $V_b = 0$ in two terminal configurations, where the contact separation ranges from 200 nm to 1,490 nm. For all devices mentioned above and in the main text, the contacts are separated by 1-2 μm. However, if we reduce the contact separation to a few hundred nanometers (270 nm), the metal contacts prevent the polarization from switching. For contact separation ($L$) above 480 nm, the transfer characteristics show similar hysteric behavior as in Fig. 1c&d and Extended Data Fig. S3a. Because $V_b$ is always grounded, as one sweeps $V_t$, both $E_\perp$ and $n_e$ change simultaneously.

13